 \documentclass[final,3p,times,authoryear]{elsarticle}
\newcommand{\ie}{{i.e.},~}
\newcommand{\eg}{{e.g.}~}

\usepackage{amsmath,amssymb,bm,siunitx,mathtools}
\usepackage{graphicx}% Include figure files
\graphicspath{{images/}}
\usepackage{microtype}
\usepackage{booktabs,multirow,tabularx} % tables

\usepackage{hyperref} % add hypertext capabilities
\usepackage{color}\definecolor{darkblue}{rgb}{0,0,0.5} % cusotm colors
\hypersetup{unicode=false, pdftoolbar=true, pdfmenubar=true, pdffitwindow=false, pdfstartview={FitH}, pdfcreator={}, pdfproducer={}, pdfkeywords={} {} {}, pdfnewwindow=true, colorlinks=true, linkcolor=red, citecolor=darkblue, filecolor=magenta, urlcolor=darkblue}

\usepackage[capitalise]{cleveref} % automatically figures out if ref is a figure, section, theorem, etc
\newcommand{\paperTitle}{Minimum-phase property of the hemodynamic response function, \\and implications for Granger Causality in fMRI}

\journal{}

\begin{document}

\begin{frontmatter}



\title{\paperTitle}

\author[Turner]{Leonardo Novelli}
\ead{leonardo.novelli@monash.edu}

\author[Sussex]{Lionel Barnett}

\author[Sussex,CIFAR2]{Anil Seth}

\author[Turner,Wellcome,CIFAR]{Adeel Razi}

\affiliation[Turner]{organization={Turner Institute for Brain and Mental Health, School of Psychological Sciences and Monash Biomedical Imaging, Monash University},%Department and Organization
            country={Australia}}
            
\affiliation[Sussex]{organization={Sussex Centre for Consciousness Science, Department of Informatics, University of Sussex},
            country={United Kingdom}}

\affiliation[CIFAR2]{organization={CIFAR Program on Brain, Mind, and Consciousness},
            city={Toronto},
            country={Canada}}
            
\affiliation[Wellcome]{organization={Wellcome Centre for Human Neuroimaging, University College London},
            country={United Kingdom}}

\affiliation[CIFAR]{organization={CIFAR Azrieli Global Scholars Program},
            city={Toronto},
            country={Canada}}

\begin{abstract}
Granger Causality (GC) is widely used in neuroimaging to estimate directed statistical dependence among brain regions using time series of brain activity measurements.
An important issue is that functional MRI (fMRI) measures brain activity indirectly via the blood-oxygen-level-dependent (BOLD) signal, which affects the temporal structure of the signals and distorts GC estimates.  
However, some notable applications of GC are not concerned with the GC magnitude but its statistical significance against a null hypothesis. 
This is the case for network inference, which aims to build a statistical model of the system based on directed relationships among its elements. 
The critical question for the viability of network inference in fMRI is whether the hemodynamic response function (HRF) and its variability across brain regions introduce spurious relationships, i.e., positive GC values between BOLD signals that are statistically significant---even if the GC between the neuronal signals is zero. 
It has previously been mathematically proven that such spurious statistical relationships are not induced if the HRF is minimum-phase, i.e., if both the HRF and its inverse are stable (producing finite responses to finite inputs). 
However, whether the HRF is minimum-phase has remained contentious. 
Here, we address this issue using multiple realistic biophysical models available in the literature and studying their transfer functions. 
We find that these models are minimum-phase for a wide range of physiologically plausible parameter values. 
Therefore, statistical testing of GC is plausible even if the HRF varies across brain regions, with the following limitations. 
First, the minimum-phase condition is violated for parameter combinations that generate an initial dip in the HRF, confirming a previous mathematical proof. 
Second, the slow sampling of the BOLD signal (in seconds) compared to the timescales of neural signal propagation (milliseconds) may still introduce spurious GC. 
\end{abstract}


%\begin{keyword}
%%% keywords here, in the form: keyword \sep keyword
%keyword one \sep keyword two
%%% PACS codes here, in the form: \PACS code \sep code
%%% MSC codes here, in the form: \MSC code \sep code
%%% or \MSC[2008] code \sep code (2000 is the default)
%\end{keyword}

\end{frontmatter}


\section{Introduction}
The hemodynamic response function (HRF) is an important part of the biophysical models used in functional magnetic resonance imaging (fMRI). 
The HRF describes the temporal evolution of the blood oxygen-level-dependent (BOLD) signal in response to neural activity, mediated by vascular changes and neurovascular coupling. 
However, most HRF models sit on opposite sides of the model complexity spectrum. 
The simple ``canonical'' HRF model assumes the same response function across the entire brain---typically, a mixture of Gamma functions whose parameters lack physical interpretation \citep{glover1999DeconvolutionImpulseResponse}. 
Such simplicity allows one to model the BOLD signal as a convolution of the neural activity with the canonical HRF, which guarantees mathematical tractability and simplifies model fitting \citep{frassle2021RegressionDynamicCausal}. 
On the opposite side of the spectrum, there are realistic models based on physiologically-informed nonlinear differential equations \citep{buxton1998DynamicsBloodFlow,friston2000NonlinearResponsesFMRI,stephan2007ComparingHemodynamicModels,havlicek2015PhysiologicallyInformedDynamic,aquino2014SpatiotemporalHemodynamicResponse,uludag2021DeterminingLaminarNeuronal}. 
Their parameters vary across brain regions and represent physiological mechanisms that regulate blood flow and oxygenation level. 

Here, we linearise two popular biophysical models \citep{stephan2007ComparingHemodynamicModels,havlicek2015PhysiologicallyInformedDynamic} to make two contributions. 
First, the linearised HRFs fill a gap in the model spectrum by balancing complexity and tractability. 
They preserve the biophysically interpretable parameters and their variation across the brain; yet, they still allow one to model the BOLD signal as a simple linear convolution, maintaining analytical tractability. 
Second, we examine the linear HRF properties and their implications for Granger Causality (GC) analysis. 

GC was born in econometrics but has developed into an established analysis tool in neuroimaging for inferring directed functional connectivity \citep{granger1969InvestigatingCausalRelations,seth2015GrangerCausalityAnalysis,cekic2018TimeFrequencyTimevarying}. 
In its simplest form, GC considers two brain activity time series (a source and a target) and measures the extent to which knowledge of the past of the source improves the prediction of the target activity, given knowledge of the past of the target. 
Exchanging the source and the target generally produces a different result, making GC a \emph{directed} measure of statistical dependence, unlike correlation or mutual information. 
For Gaussian probabilistic distributions, GC is equivalent to information-theoretic transfer entropy, licensing an interpretation in terms of information flow \citep{barnett2009GrangerCausalityTransfer,bossomaier2016IntroductionTransferEntropy}. 
Traditional estimation of GC has relied on fitting discrete-time autoregressive models to the data \citep{geweke1984MeasuresConditionalLinear}, focussing on the one-step-ahead prediction error. 
Here, we will focus on a more recent formulation of GC for continuous-time processes \citep{barnett2017DetectabilityGrangerCausality}, which enables the estimation of GC at any finite prediction horizon (i.e., time is modelled as a continuous variable). 

The application of GC to fMRI is complicated because the HRF affects the temporal structure of the signals and distorts GC estimates \citep{david2008IdentifyingNeuralDrivers,handwerker2012ContinuingChallengeUnderstanding,barnett2017DetectabilityGrangerCausality}. 
For example, measuring GC using BOLD signals may yield a different result than using invasive, direct electrical recordings of neuronal activity. 
Although this precludes the comparison of GC estimates, some applications of GC don't rely on the GC magnitude but on its statistical significance against a null hypothesis \citep{kaminski2001EvaluatingCausalRelations,cliff2021AssessingSignificanceDirected,gutknecht2021SamplingDistributionSingleregression}. 
One relevant example is network inference (or structure learning), which aims to build a statistical model of the system based on directed relationships among its elements \citep{roebroeck2005MappingDirectedInfluence,siggiridou2019EvaluationGrangerCausalitya,novelli2019LargescaleDirectedNetwork}. 
The critical question for the viability of network inference in fMRI is whether the hemodynamic response function (HRF) and its variability across brain regions introduce spurious relationships, \ie positive GC values between BOLD signals that are statistically significant---even if the GC between the neuronal signals is zero. 
It has been mathematically proven that such spurious statistical relationships are not induced if the HRF is \emph{minimum-phase} \citep{barnett2017DetectabilityGrangerCausality,solo2016StateSpaceAnalysisGrangerGeweke}. 
This property requires both the HRF and its inverse to be stable, \ie produce finite responses to finite inputs (the notion of minimum-phase filter and the necessary conditions will be defined in the Methods section). 
However, whether the HRF is minimum-phase has remained contentious \citep{solo2016StateSpaceAnalysisGrangerGeweke}. 
We address this question by studying the mathematical properties of various biophysical HRF models available in the literature.

\section{Methods}
\noindent Changes in BOLD signal depend on a variety of physiological processes and quantities \citep{buxton2012DynamicModelsBOLD}, including
\begin{itemize}
    \item vasodilatory signal ($s$)
    \item blood flow ($f$)
    \item venous volume ($v$)
    \item deoxyhemoglobin content ($q$)
\end{itemize}
In the biologically detailed models considered here, these are grouped into a vector column variable $\bm{x}=(s,f,v,q)^\intercal$ that evolves over time according to the hemodynamic state equation:
\begin{align}
    \dot{\bm{x}} &= F(\bm{x},u),
\end{align}
where $u$ is the neuronal input that initiates the hemodynamic response. 
The function $F$ varies between models, but the neuronal input generally induces vessel dilation and increases the incoming flow of oxygenated blood, leading to deoxyhemoglobin content decay. 
The resulting BOLD change ($y$) is described by a second equation:
\begin{align}
    y &= G(\bm{x}).
\end{align}
Both $F$ and $G$ are nonlinear functions.
Here, $F$ and $G$ will be taken from two important hemodynamic models \citep{stephan2007ComparingHemodynamicModels,havlicek2015PhysiologicallyInformedDynamic}. 
As we will explain in \cref{sec:minimumphaseproperty}, testing the minimum-phase condition is much simpler via the transfer function, which requires linearising the system of differential equations around their fixed point. 
Therefore, we will linearise the equations to obtain the state-space representation\footnote{Unlike typical state-space models used in GC and DCM, here $\bm{x}$ denotes the hemodynamic state variable (not the neuronal state variable) and $u$ denotes the neuronal activity (not the experimental input). The neuronal activity is the input that initiates the hemodynamic response.}
\begin{equation} \label{eq:state-space_general}
\begin{aligned}
    \dot{\bm{x}} &= A \bm{x} + B u \\
    y &= C \bm{x}.
\end{aligned}
\end{equation}
The matrices $A,B,C$ will then be used to compute the transfer function\begin{equation} \label{eq:transfer_function_generic}
\begin{aligned}
    H(s)    &= C(s I - A)^{-1} B,
\end{aligned}
\end{equation}
where $s$ is the Laplace variable \citep{oppenheim1997SignalsSystems}. 
The transfer function fully characterises the linearised system and depends on the model parameters, which have a direct biological interpretation. 
We will use it to study the BOLD impulse response (\ie the response to an ideal, instantaneous neuronal input) and its minimum-phase properties, which have important implications for GC analysis.\footnote{Studying the impulse response to an ideal, instantaneous input (a Dirac delta function) is standard practice in control theory. In this context, we are modelling a neuronal input, not an external experimental input as in task-based fMRI studies.}

\subsection{Minimum-phase property of a transfer function}
\label{sec:minimumphaseproperty}
\noindent A function is a \emph{minimum-phase} filter if it is
\begin{itemize}
    \item causal (the output only depends on past and present values of the input), 
    \item stable (the response to a finite input is always finite), 
    \item invertible, and 
    \item its inverse is also stable and causal. 
\end{itemize}
The mathematical definitions can be found in standard linear control theory textbooks, \eg \citet{oppenheim1997SignalsSystems}. 
If all these conditions are met, the inverse system is also minimum-phase. 
In other words, an inverse filter exists that is physically realisable, bounded, and can perfectly reverse the original system’s effects. 
In practice, we will use an equivalent definition that is easier to verify: the state-space model in \cref{eq:state-space_general} is minimum-phase if and only if all the zeros and poles of its transfer function have a negative real part. 
The zeros are the roots of the numerator of the transfer function, and the poles are the roots of its denominator (\ie the values that make the denominator equal to zero). 
After obtaining the transfer function via \cref{eq:transfer_function_generic}, we will find its zeros and poles to determine the parameter ranges that make the hemodynamic response function minimum-phase.

\section{Results}
\label{sec:results}

\subsection{Stephan et al. (2007) model}
\noindent The model proposed by \citet{stephan2007ComparingHemodynamicModels} builds on the Balloon-Windkessel equations and previous models \citep{buxton1998DynamicsBloodFlow,friston2000NonlinearResponsesFMRI,obata2004DiscrepanciesBOLDFlow}. 
Currently, this is the default HRF used in the SPM12 software \citep{johnashburner2020SPM12Manual}. 
It involves four hemodynamic state equations and one BOLD observation equation:
%\begin{equation}
%\begin{aligned}
%    \dot{s} &=-\gamma  (f-1)-k s+u  \\
%    \dot{f} &=s  \\
%    \dot{v} &=\frac{f-v^{1/\alpha }}{\tau }  \\
%    \dot{q} &=\frac{f \left(\frac{ 1-(1-E_0)^{\frac{1}{f}}}{E_0}\right)-q v^{\frac{1}{\alpha }-1}}{\tau }
%\end{aligned}
%\end{equation}
\begin{equation} \label{eq:stephan2007}
\begin{aligned}
    F(\bm{x}) &=
        \left(\begin{array}{c}
        -k s -\gamma (f-1)  \\
        s  \\
        \frac{f-v^{1/\alpha }}{\tau }  \\
        \frac{f \left(\frac{ 1-(1-E_0)^{\frac{1}{f}}}{E_0}\right)-q v^{\frac{1}{\alpha }-1}}{\tau }
        \end{array}\right) \\
    G(\bm{x}) &= V_0 \left(k_1 (1-q)+k_2 \left(1-\frac{q}{v}\right)+k_3 (1-v)\right)
\end{aligned}
\end{equation}
All variables and parameters are explained in \cref{tab:parameters}, which also indicates their typical values. 
A visual summary of the model is presented in \citep[Fig. 1]{stephan2007ComparingHemodynamicModels}. 
The fixed point of the system in \cref{eq:stephan2007}is $\bm{x_0}=(0,1,1,1)^\intercal$ but one can perform the change of variables $\bm{x} \rightarrow \bm{x}-\bm{x_0}$ so that the new fixed point is at the origin. 
Linearising the system around the origin yields the state-space representation introduced in \cref{eq:state-space_general}, involving the following $A,B,C$ matrices:
%\begin{equation}
%\begin{aligned}
%    \dot{s} &= -\gamma  f-k s  \\
%    \dot{f} &= s  \\
%    \dot{v} &= \frac{f}{\tau }-\frac{v}{\alpha  \tau }  \\
%    \dot{q} &= \frac{(E_0-(E_0-1) \log (1-E_0)) f}{E_0 \tau }-\frac{q}{\tau }+\frac{(\alpha -1) v}{\alpha  \tau }
%\end{aligned}
%\end{equation}
%And the linearised BOLD signal change equation
%\begin{equation}
%    y = V_0 (-k_1-k_2) q + V_0 (k_2-k_3) v
%\end{equation}
\begin{equation}
\begin{aligned}
    \dot{\bm{x}} &=
    \overbrace{\left(
    \begin{array}{cccc}
     -k & -\gamma  & 0 & 0 \\
     1 & 0 & 0 & 0 \\
     0 & \frac{1}{\tau } & -\frac{1}{\alpha  \tau } & 0 \\
     0 & \frac{E_0-(E_0-1) \log (1-E_0)}{E_0 \tau } & \frac{\alpha -1}{\alpha  \tau } & -\frac{1}{\tau } \\
    \end{array}
    \right)}^A \bm{x} + 
    \overbrace{\left(
    \begin{array}{c}
     1 \\
     0 \\
     0 \\
     0 \\
    \end{array} 
    \right)}^B u \\
    y &= \underbrace{\left(\begin{array}{cccc}
    0 & 0 & V_0 (k_2-k_3) & V_0 (-k_1-k_2)
    \end{array}\right)}_C  \bm{x}
\end{aligned}
\end{equation}
The transfer function of this state-space model is
\begin{equation}
\begin{aligned}
    H(s)    &= C(s I - A)^- B \\
            &= \frac{V_0 ((E_0-1) \log (1-E_0) (k_1+k_2) (\alpha  s \tau +1)-\alpha  E_0 (k_1+k_3) (s \tau +1))}{E_0 (s \tau +1) (\gamma +s (k+s)) (\alpha  s \tau +1)}
\end{aligned}
\end{equation}
where $s$ is the Laplace variable, and the other symbols are model parameters described in \cref{tab:parameters}.

The transfer function has four poles with negative real parts:
$\left(-\frac{1}{\alpha  \tau },-\frac{1}{\tau },\frac{1}{2} \left(-\sqrt{k^2-4 \gamma }-k\right),\frac{1}{2} \left(\sqrt{k^2-4 \gamma }-k\right)\right)$.
The only zero of the function is $\frac{-2.50598 \epsilon -3.66903}{1.67231 \tau  \epsilon -2.10961 \tau }$, which has negative real part when $\epsilon >1.27$.
Therefore, this is the minimum-phase condition.

\begin{figure}
    \centering\includegraphics[width=0.45\textwidth]{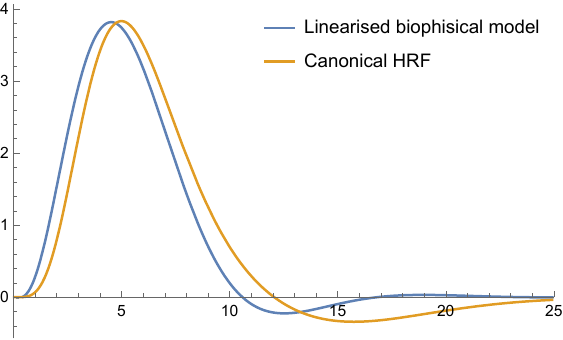}
    \caption{\label{fig:impulse_response_Stephan2007linearised_vs_canonical}
        Impulse response of the linearised biophysical model by \citet{stephan2007ComparingHemodynamicModels}, compared to the double-Gamma `canonical' response function (rescaled to achieve a peak value of $4$).
    }
\end{figure}

Finally, the hemodynamic response to an impulse (Dirac delta function $u(t)=\delta(t)$) can be obtained via the inverse Laplace transform of the transfer function. 
For the default parameter values used in SPM12, the hemodynamic response is plotted in \cref{fig:impulse_response_Stephan2007linearised_vs_canonical}. 

\subsection{Havlicek et al. (2015) model}
\noindent \citet{havlicek2015PhysiologicallyInformedDynamic} proposed the following improved model:
\begin{equation}
\begin{aligned}
    F(\bm{x}) &=
        \left(\begin{array}{c}
        -k s  \\
        \phi s -\chi (f-1)  \\
        \frac{f-\frac{f \tau +\tau_1 v^{1/\alpha }}{\tau +\tau_1}}{\tau_1}  \\
        \frac{\frac{f \left(1-(1-E_0)^{1/f}\right)}{E_0}-\frac{q \left(f \tau +\tau_1 v^{1/\alpha }\right)}{v (\tau +\tau_1)}}{\tau_1}
        \end{array}\right) \\
    G(\bm{x}) &= V_0 \left(k_1 (1-q)+k_2 \left(1-\frac{q}{v}\right)+k_3 (1-v)\right).
\end{aligned}
\end{equation}
All variables and parameters are explained in \cref{tab:parameters}, and a visual summary is presented in \citet[Fig. 1]{havlicek2017DeterminingExcitatoryInhibitory}. 
One improvement with respect to \citep{stephan2007ComparingHemodynamicModels} concerned the `balloon model`: blood vessels are still modelled as balloons that inflate and deflate with increasing or decreasing blood flow; however, the balloon initially resists a change in blood volume, which better reflects empirical findings \citep{mandeville1998DynamicFunctionalImaging}. 
Mathematically, this was achieved by adding a transient viscoelastic effect with a characteristic duration time $\tau_1$. 
Another improvement was the removal of the feedback mechanisms between the blood flow and the vasodilatory signal, which didn't agree with experiments \citep{lindauer2010NeurovascularCouplingRat,uludag2004CouplingCerebralBlood,attwell2010GlialNeuronalControl}. 

\begin{figure}
    \centering\includegraphics[width=0.8\textwidth]{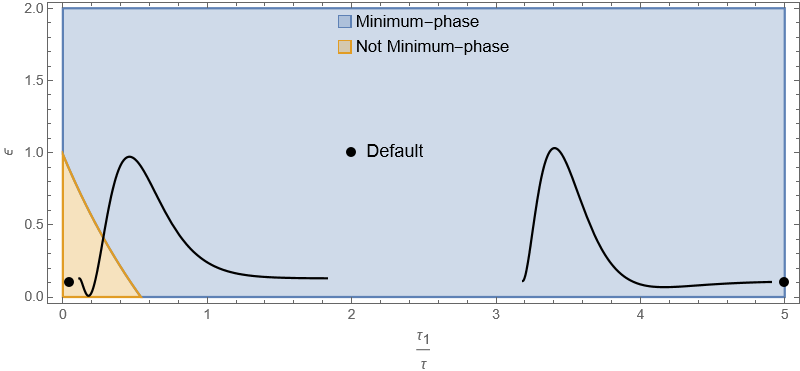}
    \caption{\label{fig:minimum-phase_region_hav2015linearised}
        The hemodynamic response function obtained by linearising the biophysical model by \citet{havlicek2011DynamicModelingNeuronal} meets the minimum-phase conditions for a wide range of parameter combinations (blue region).
        The relevant parameters are $\frac{\tau_1}{\tau}$ (the ratio of vessel viscoelastic time to mean transit time) and $\epsilon$ (the ratio of intra- to extra-vascular signal), as defined in \cref{tab:parameters}. 
        The three markers indicate the default parameters and two examples of parameter combinations that make the response minimum-phase (right) or not minimum-phase (left).
    }
\end{figure}

We use the same approach as in the previous section (a change of variable, followed by linearisation), to obtain the transfer function
\begin{equation}
\begin{aligned}
    H(s)    &= \frac{V_0 \phi  ((E_0-1) \log (1-E_0) (k_1+k_2) (\alpha  s (\tau +\tau_1)+1)-\alpha  E_0 (k_1+k_3) (s \tau_1+1))}{E_0 (s+k ) (s+\chi ) (s \tau_1+1) (\alpha  s (\tau +\tau_1)+1)} ,
\end{aligned}
\end{equation}
which has four poles with negative real parts:
$\left(-\chi ,-k ,-\frac{1}{\alpha  (\tau +\tau_1)},-\frac{1}{\tau_1}\right),\frac{1}{2} \left(\sqrt{k^2-4 \gamma }-k\right)$. 
The only zero of the function is $\frac{-3.75897 \epsilon -5.50355}{4.07905 \tau +0.588471 \tau  \epsilon +5.01694 \epsilon -6.32884}$, which has negative real part when $\tau$ or $\epsilon$ are sufficiently large. 
The minimum-phase parameter region is visualised in \cref{fig:minimum-phase_region_hav2015linearised}. 

\subsection{Canonical HRF}
\noindent The canonical HRF used in SPM12 is defined as a difference of two Gamma functions \citep{johnashburner2020SPM12Manual,lindquist2009ModelingHemodynamicResponse}:
\begin{equation}
    \frac{e^{-\frac{T t}{b_1}} \left(\frac{T t}{b_1}\right)^{a_1}}{\Gamma (a_1) T t}-\frac{e^{-\frac{T t}{b_2}} \left(\frac{T t}{b_2}\right)^{a_2}}{ \Gamma (a_2) c T t}.
\end{equation}
Unlike the previous biophysically realistic models, the parameters ($a_1,a_2,b_1,b_2,c,T$) affect the shape of the HRF (e.g., its peak value, time to peak, and duration) but lack a direct biophysical interpretation (e.g., in terms of blood flow, volume, or other physiological processes). 
The transfer function is obtained directly by taking the Laplace transform:
\begin{equation} \label{eq:canonical_HRF_transfer_function}
    \frac{c T^{a_1-1} (b_2 s+T)^{a_2}-T^{a_2-1} (b_1 s+T)^{a_1}}{c (b_1 s+T)^{a_1} (b_2 s+T)^{a_2}}
\end{equation}
With the default parameter values used in SPM12 ($a_1=6,a_2=16,b_1=16,b_2=16,c=6,T=16$), the canonical HRF has the familiar shape plotted in \cref{fig:impulse_response_Stephan2007linearised_vs_canonical}, and its transfer function is
\begin{equation}
    \frac{6 (s+1)^{10}-1}{96 (s+1)^{16}}.
\end{equation}
The only pole is at $-1$, and the $10$ zeros have negative real parts, making the default canonical HRF minimum-phase. 
Next, we vary the parameters in \cref{eq:canonical_HRF_transfer_function} to test if the minimum-phase conditions are still met. 
The zeroes and poles of the transfer function cannot be found analytically but were approximated numerically. 
Both have negative real parts for all tested parameters (default values $\pm 4$), making the canonical HRF minimum-phase over a wide range of parameters.

\section{Discussion}

By linearising two popular biophysical HRF models \citep{stephan2007ComparingHemodynamicModels,havlicek2015PhysiologicallyInformedDynamic} and studying their transfer functions, we provided the parameter ranges that make the linearised HRF models minimum-phase. 
This mathematical property is important because it ensures that the hemodynamic response doesn't introduce spurious relationships, \ie positive GC values between BOLD signals that are statistically significant even if the GC between the neuronal signals is zero. 
In addition, we studied the minimum-phase properties of the canonical HRF defined in SPM. 
This is an influential model, although its parameters lack a direct biophysical interpretation. 
Our results show that
\begin{enumerate}
    \item The canonical HRF used in SPM12 is minimum-phase for a wide range of parameter values, including the default ones used in SPM12.
    \item The model by \citet{stephan2007ComparingHemodynamicModels} involves several biophysical parameters, yet only one affects the minimum-phase property ($\epsilon$, the intra- to extra-vascular signal ratio; see \cref{tab:parameters}). 
    The condition is $\epsilon>1.27$, which is in the plausible physiological range (although the prior value used for DCM analysis is $\epsilon=1$). 
    \item The model by \citet{havlicek2015PhysiologicallyInformedDynamic} is minimum-phase for plausible parameter ranges ($\epsilon$, $\tau$, and $\tau_1$) visualised in \cref{fig:minimum-phase_region_hav2015linearised}. The fact that the minimum-phase condition is not met for all parameter values shows that it is not a trivial consequence of linearisation per se. 
\end{enumerate}
In addition, we note that the biophysical model by \citet{aquino2014SpatiotemporalHemodynamicResponse} is minimum-phase for all parameter values, based on the zeros and poles of its transfer function \citep{pang2016ResponsemodeDecompositionSpatiotemporal}. 

In summary, all HRF models considered here are minimum-phase for physiologically plausible parameter values, with some constraints. 
The parameter combinations that violate the minimum-phase condition are those that generate an initial dip in the BOLD response (see \cref{fig:minimum-phase_region_hav2015linearised}), confirming previous theoretical arguments \citep{solo2016StateSpaceAnalysisGrangerGeweke}. 
The presence of an initial dip has long been debated but is yet unresolved \citep{hillman2014CouplingMechanismSignificance,hong2018ExistenceInitialDip,taylor2018CharacterizationHemodynamicResponse}. 
Our study has two limitations. 
First, the minimum-phase property is only tested for linearised HRF models, although a more general nonlinear theory of minimum-phase systems exists \citep{byrnes1988LocalStabilizationMinimumphase}. 
Second, the slow sampling of the BOLD signal (in seconds) compared to the timescales of neural signal propagation (milliseconds) may still introduce spurious GC \citep{solo2016StateSpaceAnalysisGrangerGeweke}. 

%potentially, mention that the zero-horizon limit GC (GC rate, as opposed to finite prediction horizon) is invariant under minimum-phase HRF (although it can't be measured in practice)

Beyond the implications for GC, the linearised HRFs are useful more broadly since they balance complexity and tractability. 
They preserve the biophysically interpretable parameters and their variation across the brain; yet, they still allow one to model the BOLD signal as a simple linear convolution, maintaining analytical tractability. 
Fitting the linearised HRF models to whole-brain BOLD data can produce spatial maps of physiological processes, such as vasodilatory signal decay, transit time, and intra- to extra-vascular signal ratio. 
Future studies can use them to reveal the vascular mechanisms behind existing maps of qualitative HRF features across the brain, \eg the peak magnitude or the time-to-peak \citep{chen2023BOLDResponseMore,taylor2018CharacterizationHemodynamicResponse,handwerker2004VariationBOLDHemodynamic,lindquist2009ModelingHemodynamicResponse}. 
Furthermore, vascular changes are prominent in ageing and in neurodegenerative or psychiatric diseases \citep{tsvetanov2020SeparatingVascularNeuronal,iadecola2004NeurovascularRegulationNormal}.

\begin{table*}
\caption{
    Parameter interpretation and default values in \citep{stephan2007ComparingHemodynamicModels} and \citep{havlicek2015PhysiologicallyInformedDynamic}. 
    The values in brackets are Bayesian priors for the free parameters used in SPM12.  
    The BOLD model parameter values are for a gradient-echo sequence at $1.5$ Tesla.
    }
\label{tab:parameters}
\resizebox{\linewidth}{!}{%
\begin{tabular*}{\linewidth}{@{}llll@{}} %{l @{\extracolsep{\fill}}llll}%
\toprule
Parameters and variables & Symbol & \begin{tabular}[c]{@{}l@{}}Value \\ \citep{stephan2007ComparingHemodynamicModels}\end{tabular}  & \begin{tabular}[c]{@{}l@{}}Value \\ \citep{havlicek2015PhysiologicallyInformedDynamic}\end{tabular} \\
\midrule
Variables                            & &          \\
\midrule
Vasodilatory signal                & $s$       &     &          \\
Blood flow                           & $f$       &     &          \\
Normalised venous blood volume                  & $v$       &     &          \\
Normalised venous deoxyhemoglobin level         & $q$       &     &          \\
Hemodynamic state variable           & $\bm{x}$  & $\bm{x}=(s,f,v,q)^\intercal$ & $\bm{x}=(s,f,v,q)^\intercal$ \\
BOLD signal change (\%)              & $y$       &     &          \\
Input                                & $u$       &     &          \\
\midrule
Neurovascular coupling parameters    & &            \\
\midrule
Decay of vasodilatory signal           & $k$       & (0.64)  & 0.6      \\
Feedback regulation                  & $\gamma$  & 0.32  & -      \\
Gain of vasodilatory signal            & $\phi$    & -      & 1.5      \\
Decay of blood inflow signal         & $\chi$    & -      & 0.6      \\
\midrule
Hemodynamic model parameters         & &         \\
\midrule
Mean transit time                    & $\tau$    & (1)     & 2  \\
Vessel viscoelastic time             & $\tau_1$  & -     & 4  \\
Balloon stiffness (Grubb’s exponent) & $\alpha$  & 0.32  & 0.32     \\
Oxygen extraction factor at rest     & $E_0$     & 0.4   & 0.4      \\
\midrule
BOLD model parameters                & &         \\
\midrule
Venous blood volume fraction at rest & $V_0$     & 0.04     & 0.04        \\

\begin{tabular}[c]{@{}l@{}}Frequency offset at vessel surface for deoxygenated blood\end{tabular} & $\theta_0$ &  40.3 & 40.3 \\

\begin{tabular}[c]{@{}l@{}}Ratio of intra to extravascular signal\end{tabular} & $\epsilon$ & (1) & 0.1263-1.321 \\

\begin{tabular}[c]{@{}l@{}}Sensitivity of changes in intravascular signal\\ relaxation rate with changes in oxygen saturation\end{tabular} & $r_0$ &  25 &  15 \\

Echo time (s) & $T_\textup{E}$ & 0.04  &  0.04 \\

1st BOLD parameter                   & $k_1$     & $4.3 \theta_0 E_0 T_\textup{E}$   & $4.3 \theta_0 E_0 T_\textup{E}$ \\
2nd BOLD parameter                   & $k_2$     & $\epsilon r_0 E_0 T_\textup{E}$   & $\epsilon r_0 E_0 T_\textup{E}$ \\
3rd BOLD parameter                   & $k_3$     & $1-\epsilon$                      & $1-\epsilon$ \\
\bottomrule
\end{tabular*}%
}
\end{table*}

\section*{Author Contributions}
\noindent Leonardo Novelli: Conceptualization; Formal analysis; Investigation; Visualization; Writing --- original draft.

\noindent Lionel Barnett: Conceptualization; Writing --- review \& editing.

\noindent Anil Seth: Conceptualization; Writing --- review \& editing.

\noindent Adeel Razi: Conceptualization; Funding acquisition; Supervision; Writing --- review \& editing.

\section*{Acknowledgements}
The authors thank Pedro Valdes-Sosa and Armando Rivero for insightful discussions and input. 
This research is funded by the Australian Research Council (Refs:  DE170100128  and  DP200100757) and the Australian National Health and Medical Research Council (Investigator Grant 1194910). 
A.R. is affiliated with The Wellcome Centre for Human Neuroimaging, supported by core funding from Wellcome [203147/Z/16/Z]. 
A.R. is a CIFAR Azrieli Global Scholar in the Brain, Mind \& Consciousness Program. A.K.S and L.B. are supported by European Research Council Advanced Investigator Grant CONSCIOUS (to A.K.S); grant number 101019254.

%\appendix

%\section{Sample Appendix Section}
%\label{app:sample}
%

\bibliographystyle{elsarticle-harv} 
\bibliography{bibliography}





% \end{thebibliography}
\end{document}